# Catalogs of solar wind types and their role in solar-terrestrial physics

Irina G. Lodkina[*1], Yuri I. Yermolaev[*1], Alexander A. Khokhlachev[1]
[1]Space Research Institute, Russian Academy of Sciences, 117997 Moscow, Russia,
irina-priem@mail.ru, yermol@iki.rssi.ru

## Abstract

The response of the magnetosphere to interplanetary drivers depends on their type. The reliability of their identification affects the conclusions of the analysis of connections between the solar wind and the magnetosphere. In this work, we analyze the list of moderate and strong geomagnetic storms and their interplanetary sources for the period 2009 - 2019, presented in the work of Qiu et al. It is shown that some of the events in this list were identified incorrectly, and their interpretation differs in ~20% of cases from our catalog by Yermolaev et al. (http://www.iki.rssi.ru/pub/omni/) for types of solar wind Sheath, ICME and CIR, and in ~28% of cases from the Richardson and Cane catalog for ICME. Using the unadjusted list of Qiu et al. may lead to incorrect identification of interplanetary drivers of magnetic storms and erroneous conclusions. It is recommended to use the classification of interplanetary events from catalogs of events accepted by the scientific community as reference ones.

## 1. Introduction

Solar-terrestrial physics includes the study of the totality of all possible interactions of helio- and geophysical phenomena [1,2]. The discovery of the connection between magnetospheric disturbances and the appearance of the southern (southward) component of the interplanetary magnetic field (IMF) was one of the fundamental results in solar-terrestrial physics, which was obtained at the very beginning of the space era [3-7] and determined the main direction of research for subsequent decades. Since in the stationary solar wind (SW) the IMF lies in the ecliptic plane (i.e., there is no IMF component normal to the ecliptic plane), the stationary solar wind does not lead to disturbances of the magnetosphere, and all magnetospheric disturbances are associated with disturbed types of solar wind (so called interplanetary drivers: (1) a coronal mass ejection body (ICME, including MC magnetic clouds and Ejecta), (2) a compression region in front of the fast ICME (sheath), and (3) a compression region in front of the fast coronal hole outflow (CIR) [8-12].

Subsequent studies showed that the response of the magnetosphere to the southward component of the IMF differs in different types of drivers, i.e. the response of the magnetosphere additionally depends on the type of driver (see, for example, [13-23] and references therein), which is (1) the efficiency of generating magnetic storms according to the "output/input" criterion, where "input" refers to the parameter of the interplanetary driver, and under the “exit” is the geomagnetic index, the highest is observed in sheath, and the lowest in Ejecta; (2) both types of compression, CIR and sheath, are similar to each other, but the compression ratio in sheath is usually achieved higher; (3) both types of ICME, MS and Ejecta, have similar parameters, but in MS a higher MMF is observed. The above property of solar-terrestrial connections makes it very promising to study the response of the magnetosphere to variations in interplanetary conditions in various types of solar wind, since it allows us to study the response of the magnetosphere for several sets of specific conditions characteristic of disturbed types of solar wind. In addition to problems of studying the generation of various magnetospheric-ionospheric disturbances of the Earth by the solar wind, identification of solar wind types is also widely used in problems in solar physics and heliosphere physics. During the space era, a large amount of statistical material has been collected on measurements of solar wind parameters near the Earth (see http://omniweb.gsfc.nasa.gov and article [24]) and a large number of studies have been published.

As an analysis of a large number of publications has shown, this approach has serious methodological problems. Most works, as a rule, use a fairly limited set of events without a sufficiently detailed description of the nomenclature and procedures for identifying solar wind types. This often does not allow one to verify the analyzed sets of interplanetary drivers and draw reliable conclusions, confirmed by several studies. Some of the work uses incorrect data sets or incorrect methods for analyzing them. Some overview of the most common errors with examples is contained in our article [23] and in the references given therein. Here we will only briefly mention that in addition to those works in which it is obvious that the authors use incorrect, different from generally accepted, criteria for identifying interplanetary drivers (for example, [25]), there is a class of works in which such errors are not obvious. They are usually related, either

(1) using incorrect data from previously published works (for example, work [26], which uses incorrect data on the identification of SW types from the list of Shen et al. [27] or work [28], which used the results of incorrect identification contained in the list [29] or

(2) considering the so-called “CME-driven” event. As shown above, a CME observed near the Sun in Earth’s orbit can generate 2 geoeffective events: an ejection (Ejecta/MC) and a compression region in front of it (sheath), which have different characteristics [11, 23, 30] and, as a consequence, , different response of the magnetosphere [19, 23]. When considering CME-driven events (without their selection for Ejecta/MC emission and sheath compression region), a certain “average phenomenon” that does not exist in nature with an unknown contribution from 2 different drivers is studied. The results obtained in this way, in our opinion, have no physical meaning, since they characterize only a certain “average” course of parameters for a specific set of different events, which may differ from any similar set due to different contributions from Ejecta/MC and sheath, i.e. .e. the results depend on the sample of events used and are random in nature.

The above shows that the methodological errors of this approach are not related to the object of research, but are exclusively subjective in nature, since the authors of the studies (and reviewers of the relevant articles) do not pay enough attention to methodological issues of research, as a result of which the number of publications in which there are no scientific conclusions. As we have already indicated [23], this problem is easily solved by introducing into scientific circulation catalogs of SW events, accepted by the scientific community as reference ones. We saw no objections to this proposal. However, there is currently no organizational and financial support for this idea. At the same time, there are at least 2 catalogs that cover several solar cycles in time and have more than $10^2$ references in the literature, which could form the basis of such a catalog agreed upon by the scientific community.

1. Catalog of large-scale types of solar wind, which has been developed for more than a quarter of a century at the IKI RAS (http://www.iki.rssi.ru/pub/omni/ [31]). This catalog includes 1-hour data from 1976 through 3-6 months of OMNI+ database updates and intervals of the following 7 large-scale (>$10^6$ km) solar wind types:

quasi-stationary types (1) Heliospheric current sheet, HCS, (2) slow flows from the region of coronal streamers, Slow, (3) fast flows from the region of coronal holes, Fast, and perturbed types (4) compression regions between the slow and fast types of flow - corotating interaction regions (CIR), (5) areas of compression between the slow type of flow and rapid manifestations of CME (ICME), Sheath, and (6, 7) 2 variants of ICME, - Ejecta and magnetic cloud ( MC), which are distinguished by a higher and more regular interplanetary magnetic field (IMF) in the MC compared to Ejecta. The classification of solar wind phenomena used is generally accepted (for more details, see [23]); the method for identifying types is described in detail in [31].

2. Catalog of Near-Earth Interplanetary Coronal Mass Ejections I.G. Richardson and H.V. Cane (https://izw1.caltech.edu/ACE/ASC/DATA/level3/icmetable2.htm), which includes ICME measurement data since 1996. Due to the fact that this catalog, in addition to interplanetary data on SW, includes solar information on CME obtained using coronagraph measurements, this catalog is very popular for problems in the physics of the Sun and the heliosphere, but it has limited application for problems in geophysics. This is due to the fact that this catalog does not include data on the important interplanetary driver of magnetospheric disturbances - CIR. In addition, the catalog considers a composite CME-driven phenomenon, including shock wave, Ejecta/MC and sheath, as an ICME. As noted above, to obtain a rigorous scientific result in geophysics, it is necessary to carry out additional selection of such “ICMEs” into their two geoeffective parts: Ejecta/MC and sheath. For individual sets of events, we carried out such work on selecting phenomena from the Richardson and Cane catalog and comparing them with the results of our catalog (see, for example, works [23, 30]) and references in them, and the agreement between the data of both catalogs is about 13%. It should be noted that we did not set a special task for a systematic comparison of the two catalogs, and such work was carried out when assessing the reliability of the identification of interplanetary drivers and conclusions in some published studies, in which the authors of these works neglected a detailed description of methodological issues. Below, as an illustration, we present one of the latest similar works, compare it with the data of the two mentioned catalogs and give a conclusion about the reliability of driver identification and the conclusions made by the authors.

In a recent work [32], the authors study the excitation on Earth of 149 magnetic storms (MS) with Dstmin ≤ 50 nT by various interplanetary drivers over the period of the 24th solar cycle (2009–2019). This work could be of scientific interest if the following shortcomings were absent. As follows from the text of the article, the authors are aware of the presence of two widely used databases of solar wind types, the catalog of Richardson, I.G., and H.V. Cane and our catalog at IKI RAS. Although the specified databases contained data on the types of drivers for the specified intervals at the time of preparation of the manuscript, the authors chose not to use the published results in the specified databases, but to conduct their own identification of drivers. There are no results of a quantitative comparison of the results of our own analysis with the results of other similar studies in [32], and the authors limited themselves to the differences with the above bases by saying “Although these inconsistencies are ineluctable but they put little to no effect on our findings"), which do not contain quantitative characteristics. The work and the reference to the methodology [33] contained in the work describe the qualitative criteria of ICME (only the body of the CME, sheath is not mentioned at all in the article [33]), which do not allow reproducing the analysis of the source data and verifying the list of events. The purpose of this work is to compare the results of identifying interplanetary drivers from [32] with the identification of the same events using the databases of Richardson, I.G., and H.V. Cane and IKI RAS to obtain quantitative estimates of the degree of agreement between the identification of drivers of these databases and the reliability of the database data used in the article [32].

## 2. Methodology

In this work, we analyze 3 different data sources that provide the results of identifying the types of SWs and interplanetary drivers for the same time interval (2009-2019) that is used in [32].

(1) Results of identification of interplanetary drivers of magnetic storms (MS) with Dst < -50 nT in the period 2009-2019 in the list of Qiu et al (see Table 1 in [32],

(2) Data from the catalog of Yermolaev et al., [31] of interplanetary phenomena, which includes the results of identification of large-scale types of SW Sheath, ICME (separately MC or Ejecta) and CIR for the period 1976-2022 and which is freely available on the website of the IKI RAS at http://www.iki.rssi.ru/pub/omni/ (or ftp://ftp.iki.rssi.ru/pub/omni/ via ftp protocol).

(3) Richardson and Cane catalog data [34] for the “ICME“ phenomena (https://izw1.caltech.edu/ACE/ASC/DATA/level3/icmetable2.htm), including both the CME body and the region squeezing Sheath in front of him.

Our methodology for identifying SW phenomena is based on experimental data that showed that the Sheath and CIR compression regions are characterized by an increase in velocity, density, temperature and plasma β-parameter, while in ICME these parameters decrease. The detailed procedure for analyzing data from the OMNI and OMNI2 databases (see http://omniweb.gsfc.nasa.gov and article [24] and the identification of 8 types of SVs is described in [31]. For analysis, we use intervals for the following types and subtypes:

- IS/Sheath/Ejecta – Sheath intervals, accompanied by a preceding shock wave and a subsequent Ejecta interval,
- Sheath/Ejecta - Sheath intervals without a preceding shock wave and with a subsequent Ejecta interval,
- Ejecta – Ejecta intervals without a preceding Sheath interval
- IS/Sheath/MC - Sheath intervals, accompanied by a preceding shock wave and a subsequent MC interval,
- Sheath/MC - Sheath intervals without a preceding shock wave and with a subsequent MC interval,
- MC - MC intervals without a preceding Sheath interval,
- CIR - CIR intervals without a preceding shock wave
- IS/CIR - CIR intervals from the previous shock wave.

Annual lists of intervals of various types of SW in the form of text files containing the start and end times of intervals of various types of SW, located in the catalog http://www iki.rssi.ru/pub/omni/catalog/.

The results are presented in graphical and digital form and are freely available on the IKI RAS website http://www.iki.rssi.ru/pub/omni/. The technique described in our previous article [30] was also used. When using this method, the time course of the parameters for the selected event was compared with the course of these parameters averaged over a set of one or another type of SW over 25 years [36, 35], i.e. Not only the values of the parameters were compared, but also their dynamics over time.

We compared the results of identifying the types of SW events from the list of Qiu et al with the identification in the Yermolaev et al. catalog and the Richardson and Cane catalog. Since the Richardson and Cane catalog does not contain data on CIR, only data from Qiu et al, classified by the authors as ICME and Sheath, were analyzed for comparison with this catalog. The results are discussed in detail in the next section. Also in the next section are some examples with measurement data from the Yermolaev et al. catalog for intervals for which the interpretation differs in the list of Qiu et al.3.

**Results**.

In this section, we first (in Section 3.1) present the statistical results of comparing the identification of types of SW events in the list of Qiu et al., in the catalog of Yermolaev et al., and the Richardson and Cane catalog. We then (in Section 3.2) take a closer look at some of the events that have different identities across these three directories.

### 3.1. Directory comparison statistics

In column 2 of Table. Table 1 presents 149 moderate and strong geomagnetic storms and their interplanetary sources for the period 2009–2019, generated by various types of SW (ICME, CIR, SH - Sheath, as well as the types of SW Complex and Unknown) according to the list of Qiu et al. The next two columns show a comparison with the catalog of Yermolaev et al. for the same intervals: in column 3 "+" and "-" indicate a match or mismatch of drivers, column 4 indicates the identification of drivers in the catalog of Yermolaev et al. (MC - magnetic cloud , MCsh is the magnetic cloud after the Sheath compression region, EJ is the Ejecta piston, EJsh is the Ejecta piston after the Sheath compression region, SHej is the compression region in front of the fast Ejecta, SHmc is the compression region in front of the fast MC, SW is the unperturbed SW). The 5th column shows the presence ("+") or absence ("-") of an event from the list of Qiu et al. in the Richardson and Cane catalog, which includes only events of the ICME and Sheath type, and the 6th column is a match of event identification in the Yermolaev et al. and Richardson and Cane catalogs.

Our analysis of events, presented in Table. 1 shows a noticeable discrepancy between the identification of drivers and the catalog data of Yermolaev et al., which amounts to 19.5% of cases (out of 149 events, 29 events have discrepancies in identification). Table 2 in this work contains links to the drawings of the Yermolaev et al. catalog for 29 events that have differences in the identification of the Qiu et al list from Table. 1., and uses the same driver type designations.

Of the 81 events listed by Qiu et al that are labeled by the authors as ICME or Sheath,
23 events are missing from the Richardson and Cane catalogue, i.e. in ~28% of cases in this catalog they are not identified as Sheath or ICME. Since the nomenclature of SW types in the Yermolaev et al. and Richardson and Cane catalogs differs, the differences in identification according to these catalogs can be estimated approximately as 13%.
Thus, analysis of data from three catalogs of SW types shows that a significant number of identified events from the list of Qiu et al differ from both the Yermolaev et al. catalog and the Richardson and Cane catalog.

Results of comparison of identifications of some events Table. 1 presented in various catalogs are discussed in detail below.

### 3.2. Interpretation of events that have discrepancies in the list of events Qiu et al.

55 events out of 149 in the list of Qiu et al are marked by the authors as type CB CIR (Table 1). According to the catalog by Yermolaev et al., 10 events have differences:

- 5 events numbered 11, 16, 109, 115, 134 are identified as ICME with preceding Sheath events,
- 3 events numbered 66, 89, 143 are identified as Sheath followed by ICME,
- 2 events 137 and 149 from the list of Qiu et al fall on the SW (unperturbed SW).

Let's consider the event marked in the list by Qiu et al as CIR number 134 with the time 31.VIII.2017 12.00 UT (hereinafter world time). This event is the minimum of the MS with Dst = -50 nT, identified in our catalog as Ejecta with a preceding Sheath with a shock (see Fig. 1 and Table 2).

In Fig. 1 (and further in Fig. 1-5) shows the time course of the parameters located on 7 panels (from top to bottom), with the boundaries of the event highlighted by vertical lines:

1st panel: β-parameter – ratio of thermal pressure to magnetic pressure (filled circles), T/Texp – relative temperature of SW protons (crosses),

NkT is the thermal pressure of SW (unfilled diamonds), Na/Np is the ratio of the concentrations of alpha particles to protons (unfilled triangles).

Panel 2: B – IMF value (filled circles), Bz – IMF component (crosses), Bx – IMF component (unfilled triangles), DB – gradient (increment) of the IMF value over an interval of 6 hours (unfilled diamonds).

Panel 3: T – plasma (proton) temperature (filled circles), Texp – expected average temperature at the measured SW velocity (crosses).

Panel 4: N - SW concentration (filled circles), $mnV^2$ (indicated as $nV^2$ in the figure) – kinetic pressure of SW (crosses), DN – gradient (increment) of plasma concentration over an interval of 6 hours (unfilled diamonds)

Panel 5 - V – SW speed (filled circles), DV6 – gradient (increment) of the speed value over an interval of 6 hours (crosses).

Panel 6: Kp – index (solid line), Ey – electric field (crosses).

Panel 7: Dst – index (black line), Dst* – adjusted Dst index (crosses).

In the above interval, an increase in V is observed, accompanied by an increase in T, N, plasma β-parameter, B, NkT and $mnV^2$ ($nV^2$). This behavior of the parameters is typical for the compression regions

CIR and Sheath [37,38], but since after this region the SW Ejecta type is observed (decrease in parameter values: the magnitude of the magnetic role B, the thermal value of the solar wind NkT, plasma density N, velocities V), then this event is identified as Sheath. The MB minimum with a minimum of Dst = -50 nT and a date of 31.VIII.2017 12.00 is located within this Ejecta interval.

Similarly, events 11, 16, 109 and 115, according to our catalog, are observed within ICME intervals, with a preceding Sheath and with a shock wave (Table 2), and not CIR intervals.

Events 66, 89, and 143 from the list of Qui et al (Table 1), marked as CIR, are identified in our catalog as Sheath followed by ICME.

Let us consider in detail event 66 with the minimum time of MB on November 9, 2013, 09.00 and Dst = –80 nT. In Fig. Figure 2 highlights the interval in which the minimum MB falls on Sheath (accompanied by an increase in the values of T, N, plasma β-parameter, B, NkT and $mnV^2$ ($nV^2$) as noted above, this behavior of parameters is typical for this type of SW and is observed before Ejecta ( decrease in the values of parameters B, NkT, V, N)).

For events 66, 89, 143 in Table. 2 contains information about the figures in the catalog of Yermolaev et al.

Events 137 and 149 from the list of Qiu et al (Table 1) according to our catalog fall on the SW (unperturbed SW). The CIR interval for event 137 ended 50 hours before the MS minimum (10/14/2017 06:00 with Dst = –52 nT), and for event 149 17 hours before the MS minimum (09/01/2019–09 07:00 with Dst = –52 nT ). Links to figures in the catalog by Yermolaev et al. are presented in Table. 2.

The Richardson and Cane catalog identifies events 16, 109, 66 as ICME rather than CIR.

Event 58, according to the interpretation of Qiu et al., falls on Sheath, and according to the catalog by Yermolaev et al., on Ejecta (see Fig. 3 and Table 2), since this interval is characterized by a decreasing velocity, a moderate magnitude of the magnetic field and low values of temperature, concentration and beta parameter. It should be noted that the Ejecta region was observed approximately 24 hours after the arrival of the interplanetary shock wave, but the question is whether this shock wave is associated with the Ejecta (i.e., whether the Sheath region precedes the Ejecta as observed in Fig. 4) or no, it requires additional research and is beyond the scope of this article.

Events 39, 45, 63, 135 of the list by Qiu et al, according to the interpretation of the authors of the list, are in the ICME, and according to the catalog of Yermolaev et al. they are observed in the Sheath interval.

Let us consider in detail event 39 with a minimum at 11.00 on September 3, 2012 and Dst = -69 nT. According to Fig. 4 (see Table 2) the MS minimum falls on Sheath, which is characterized by an increase in velocity V on the shock wave at the beginning of the interval, accompanied.

## 4. Discussion and conclusions

In this work, we examined the results of identifying all interplanetary drivers of magnetic storms presented in the list of Qiu et al. In order to assess the reliability of the proposed identification of interplanetary storm drivers, we compared these results with the identification of these intervals in our Yermolaev et al. catalog and the data from the Richardson and Cane catalog. In Section 3, we examined events that, according to our methodology, were identified incorrectly. In Table. 1 and 2, the results of the analysis are presented. Of all the events listed by Qiu et al in Table. 1 of this work, and comparing them with data from the catalog of Yermolaev et al. and Richardson and Cane, we draw the following conclusion.

(1) Of the 149 events listed by Qiu et al according to the catalog by Yermolaev et al., 29 have differences in identification and only 9 of the 29 events are in the Richardson and Cane catalog.

(2) In Qiu et al's list of 81 events labeled ICME or Sheath, 23 events are missing from the Richardson and Cane catalogue.

(3) Under certain assumptions, the differences in driver identification in the Yermolaev et al. catalog and in the Richardson and Cane catalog can be estimated as 20 out of 149 events. This value is consistent with the previously obtained estimate of differences of about 13% for these catalogs.

Thus, the identification of interplanetary driver types in the list of Qiu et al differs from the identification in the Yermolaev et al. catalog in 19% and the Richardson and Cane catalog in ~28% of events. In our opinion, the differences in identifying drivers are significant, and using the results of the list of Qiu et al without explanations from the authors can lead to false conclusions in problems studying the connections between interplanetary and magnetospheric phenomena. As we have already indicated [23], this problem is easily eliminated if we use the identification of interplanetary events from catalogs of events accepted by the scientific community as reference ones.

## Acknowledgments

The authors are grateful to the creators of the OMNI database and the ICME catalog, Richardson and Cane, for the opportunity to use the data. OMNI data obtained from GSFC/SPDF OMNIWeb interface at

http://omniweb.gsfc.nasa.gov, ICME catalog data at http://www.srl.caltech.edu/ACE/ASC/DATA/level3/icmetable2 .htm/. The work was carried out with financial support from the Russian Science Foundation, grant No. 22-12-00227.

## References

1. Zelenyi L.M., Veselovsky I.S. Plasma heliogeophysics. Moscow. M.: Fiz-matlite, 2008. T. 1. 672 c.; T. 2. 560 c.
2. Zelenyi L.M., Petrukovich A.A., Veselovsky I.S. Modern achievements in plasma heliogeophysics. Moscow. M.: IKI RAN, 2016. — 672 c. — ISBN 978-5-00015-011-5.
3. Dungey J.W. Interplanetary Magnetic Field and the Auroral Zones // Phys. Rev. Lett. 1961, 6, 47–48.
4. Fairfield D.H. Cahill L.J. The transition region magnetic field and polar magnetic disturbances // J. Geophys. Res. 1966, 71, 155–169.
5. Rostoker G., Falthammar C.-G. Relationship between changes in the interplanetary magnetic field and variations in the magnetic field at the Earth's surface // J. Geophys. Res. 1967, 72, 5853–5863.
6. Russell C.T., McPherron R.L., Burton R.K. On the cause of magnetic storms // J. Geophys. Res. 1974, 79, 1105–1109.
7. Burton R.K., McPherron R.L., Russell C.T. An empirical relationship between interplanetary conditions and Dst // J. Geophys. Res. 1975, 80, 4204–4214
8. Tsurutani B.T., Gonzalez W.D. The interplanetary Causes of Magnetic Storms: A Review. In Magnetic Storms; Mon S., Tsurutani B.T., Gonzalez W.D., Kamide Y., Eds.; // American Geophysical Union Press: Washington, DC, USA, 1997; Volume 98, p. 77.
9. Gonzalez W.D., Tsurutani B.T., Clua de Gonzalez A.L. Interplanetary origin of geomagnetic storms // Space Sci. Rev. 1999, 88, 529–562.
10. Yermolaev Y.I., Yermolaev M.Y., Zastenker G.N., Zelenyi L.M., Petrukovich A.A., Sauvaud J.A. Statistical studies of geomagnetic storm dependencies on solar and interplanetary events: A review // Planet. Space Sci. 2005, 53, 189–196.
11. Yermolaev Y.I., and Yermolaev M.Y. Solar and Interplanetary Sources of Geomagnetic Storms: Space Weather Aspects // Izvestiya, Atmospheric and Oceanic Physics, 2010, Vol. 46, No. 7, pp. 799–819. 2010.
12. Temmer M. Space weather: The solar perspective // Living Rev. Sol. Phys. 2021, 18, 4.
13. Eselevich V.G., Fainshtein V.G. An investigation of the relationship between the magnetic storm Dst indexes and different types of solar wind streams // Ann. Geophys. 1993, 11, 678–684.
14. Huttunen K.E.J., Koskinen H.E.J., Schwenn, R. Variability of magnetospheric storms driven by different solar wind perturbations // J. Geophys. Res. 2002, 107
15. Borovsky J.E., Denton M.H. Differences between CME-driven storms and CIR-driven storms // J. Geophys. Res. 2006, 111
16. Pulkkinen T.I., Partamies N., Huttunen K.E.J., Reeves G.D., Koskinen H.E.J. Differences in geomagnetic storms driven by magnetic clouds and ICME sheath regions // Geophys. Res. Lett. 2007, 34, L02105.
17. Yermolaev Y.I., Nikolaeva N.S., Lodkina I.G., Yermolaev M.Y. Relative occurrence rate and geoeffectiveness of large-scale types of the solar wind // Cosm. Res. 2010a, 48, 1–30.
18. Yermolaev Y.I., Nikolaeva N.S., Lodkina I.G., Yermolaev M.Y. Specific interplanetary conditions for CIR-induced, Sheath-induced, and ICME-induced geomagnetic storms obtained by double superposed epoch analysis // Ann. Geophys. 2010b, 28, 2177–2186
19. Yermolaev Y.I., Nikolaeva N.S., Lodkina I.G., Yermolaev M.Y. Geoeffectiveness and efficiency of CIR, sheath, and ICME in generation of magnetic storms // J. Geophys. Res. 2012, 117, A00L007
20. Nikolaeva N., Yermolaev Y., Lodkina I. Predicted dependence of the cross polar cap potential saturation on the type of solar wind stream // Adv. Space Res. 2015, 56, 1366–1373.

21. Despirak I.V., Lyubchich A.A., Kleimenova N.G. SolarWind Streams of Different Types and High-Latitude Substorms // Geomagn. Aeron. 2019, 59, 1–6
22. Dremukhina L.A., Yermolaev Y.I., Lodkina I.G. Dynamics of Interplanetary Parameters and Geomagnetic Indices during Magnetic Storms Induced by Different Types of Solar Wind // Geomagn. Aeron. 2019, 59, 639–650
23. Yermolaev Y.I., Lodkina I.G., Dremukhina L.A., Yermolaev M.Y., Khokhlachev A.A. What Solar–Terrestrial Link Researchers Should Know about Interplanetary Drivers. 21 //Universe 2021, 7, 138. https://doi.org/10.3390/universe7050138
24. King J.H. and Papitashvili N.E. Solar wind spatial scales in and comparisons of hourly wind and ACE plasma and magnetic field data // J. Geophys. Res., 2004, vol. 110, no. A2, A02209. https://doi.org/10.1029/2004JA010804
25. Hutchinson J.A., Wright D.M., Milan S.E. Geomagnetic storms over the last solar cycle: A superposed epoch analysis // J. Geophys. Res. 2011. V. 116. A09211. https://doi.org/10.1029/2011JA016463
26. Pandya M., Veenadhar B., Ebihara Y. et al. Variation of Radiation belt electron flux during CME and CIR driven geomagnetic storms: Van Allen Probes observations // JGR Space Physics. 2019. DOI:10.1029/2019JA026771
27. Shen X.-C., Hudson M. K., Jaynes A. et al. Statistical study of the storm time radiation belt evolution during Van Allen Probes era: CME- versus CIR-driven storms. // J. Geophys. Res. Space Physics. 2017. V. 122. P. 8327–8339. doi:10.1002/2017JA024100
28. Ogawa Y., Seki K., Keika K., Ebihara Y. Characteristics of CME- and CIR-driven ion upflows in the polar ionosphere // JGR Space Physics. 2019. V. 124. P. 3637– 3649]
29. Kataoka R. and Miyoshi Y. Flux enhancement of radiation belt electrons during geomagnetic storms driven by coronal mass ejections and corotating interaction regions // Space Weather, 2006, vol. 4, S09004. https://doi.org/10.1029/2005SW000211
30. Yermolaev, Yu.I., Lodkina, I.G., Nikolaeva, N.S., et al., Some problems of identifying types of large-scale solar wind and their role in the physics of the magnetosphere // Cosmic Res., 2017, vol. 55, no. 3, pp. 178–189
31. Yermolaev Yu. I., Nikolaeva N.S., Lodkina I.G., Yermolaev M.Y. Catalog of Large-Scale Solar Wind Phenomena during 1976-2000 // Cosm. Res. 2009, 47, 81–94.
32. Qiu S., Zhang Z., Yousof H., Soon W., Jia M., Tang W., Dou,X. The interplanetary origins of geomagnetic storm with Dstmin ≤ 50nT during solar cycle 24 (2009–2019) // Advances in Space Research, 2022. https://doi.org/10.1016/j.asr.2022.06.025
33. Shen C., Y. Wang Z., Pan B., Miao P. Ye, and S. Wang (2014). Full-halo coronal mass ejections: Arrival at the Earth // J. Geophys. Res. Space Physics, 119, 5107–5116, doi:10.1002/2014JA020001
34. Richardson, I.G., Cane, H.V. Near-Earth Interplanetary Coronal Mass Ejections During// Solar Cycle 23 (1996 – 2009): Catalog and Summary of Properties // Sol. Phys., 2010. Vol. 264: 189, https://doi.org/10.1007/s11207-010-9568-6
35. Yermolaev, Y.I., Lodkina, I.G., Nikolaeva, N.S., and Yermolaev, M.Y., Dynamics of large-scale solar-wind streams obtained by the double superposed epoch analysis: 2. Comparisons of CIR vs. Sheath and MC vs. Ejecta // Sol. Phys., 2016, vol. 292. https://doi.org/10.1007/s11207-017-1205-1
36. Yermolaev Y.I., Lodkina I.G., Nikolaeva N.S., and Yermolaev, M.Y. Dynamics of large-scale solar wind streams obtained by the double superposed epoch analysis // J. Geophys. Res.: Space Phys., 2015, vol. 120, no. 9, pp. 7094–7106. https://doi.org/10.1002/2015JA021274
37. Nikolaeva N., Yermolaev Y., Lodkina I. Modeling the time behavior of the Dst index during the main phase of magnetic storms generated by various types of solar wind // Adv. Space Res. 2013, 6, 401–412. https://doi.org/10.1134/S0010952513060038
38. Seki K., Keika K., Ebihara Y. Characteristics of CME- and CIR-driven ion upflows in the polar ionosphere // JGR Space Physics, 2019, Vol. 124, P. 3637-3649, https://doi.org/10.1029/2018JA025870

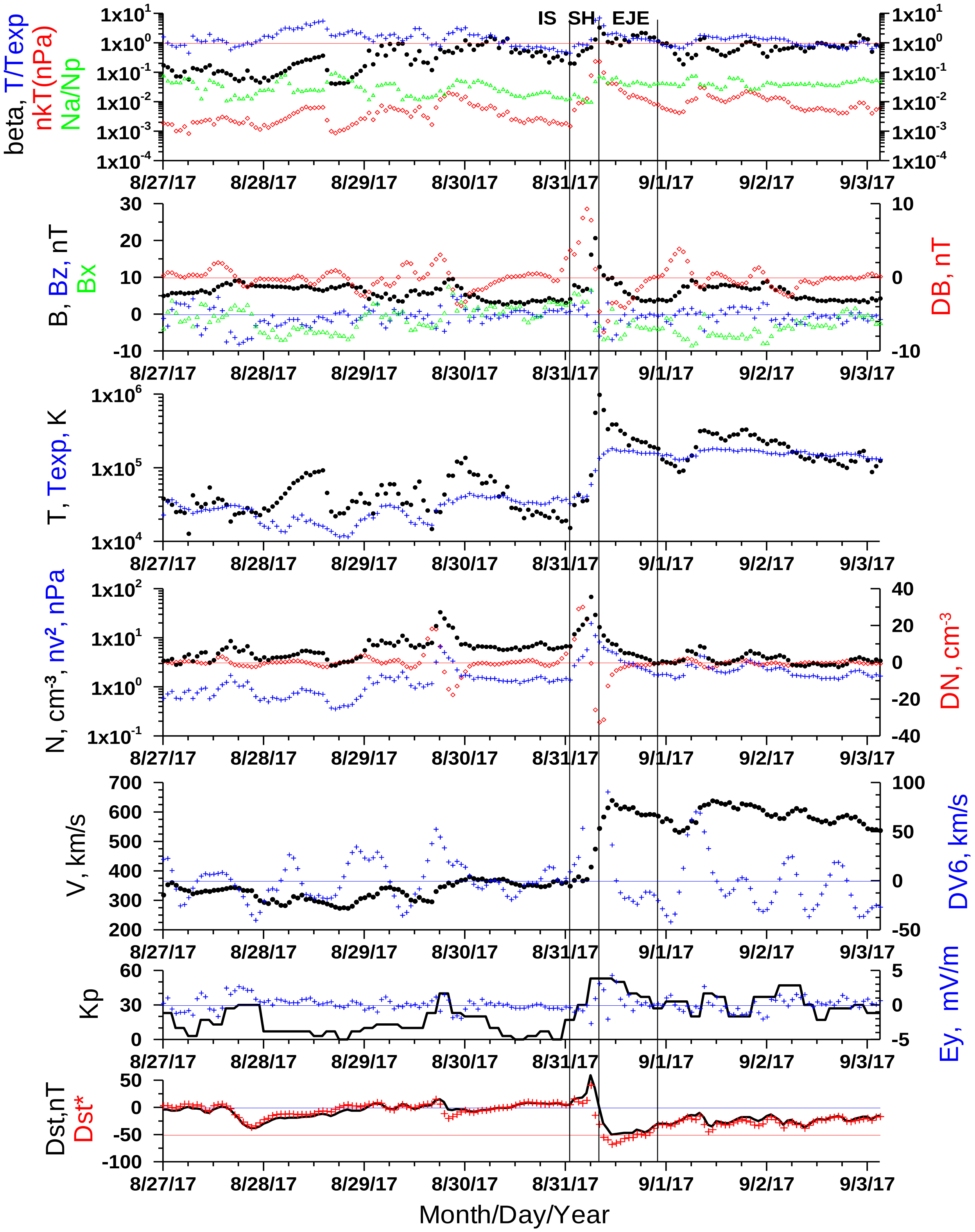

*Fig. 1. Time course of parameters of the interplanetary medium and magnetospheric indices from August 27 to March 3, 2017. (see description in text). Event 134 of the list by Qiu et al with a minimum on August 31, 2017 12.00 with Dst = –50 nT according to the catalog of* Yermolaev *et al. falls within the Ejecta interval.*

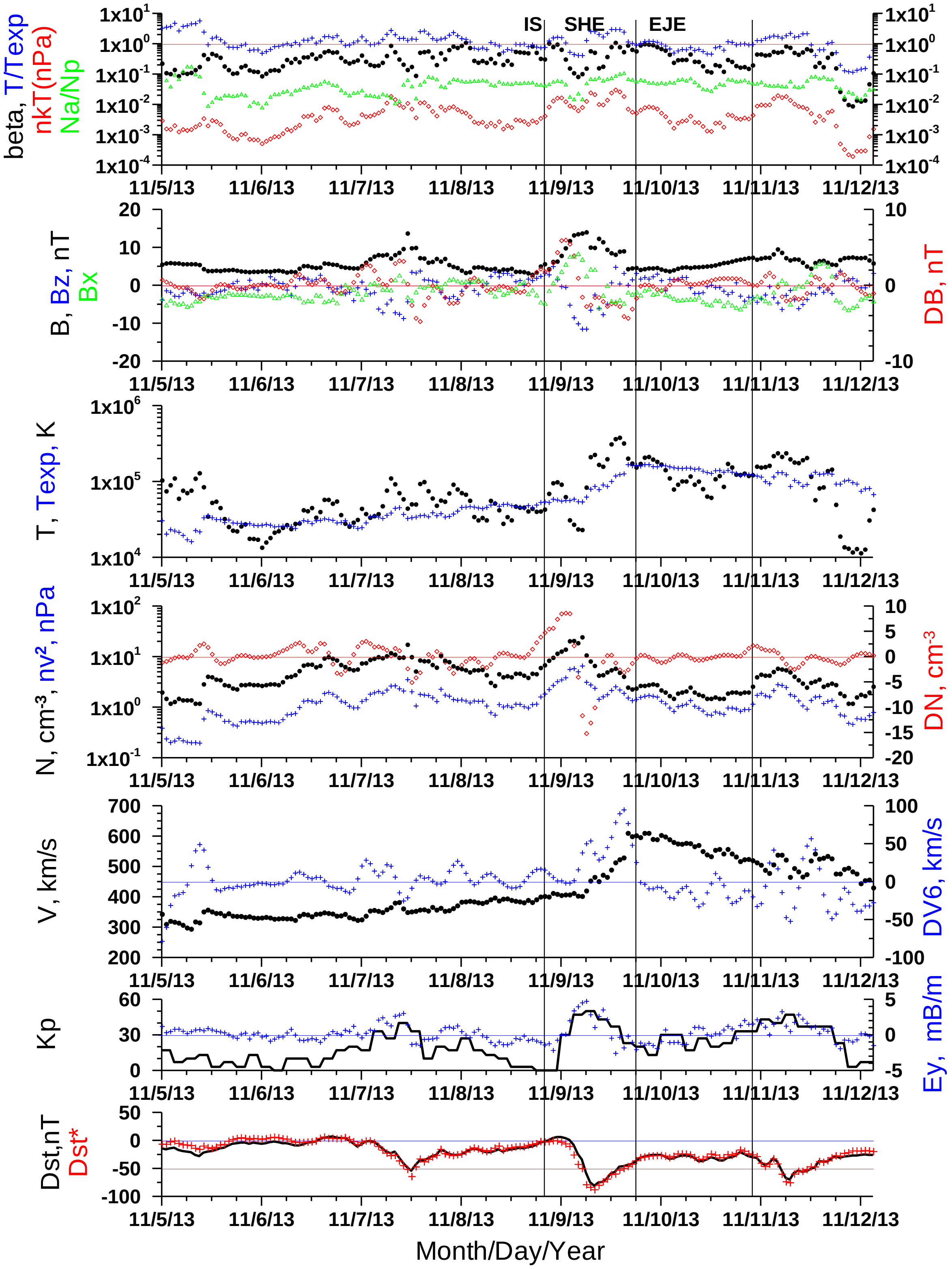

*Fig. 2. Same as fig. 1. Event 66 of the list by Qiu et al with a minimum at 09.11.2013 09.00 with Dst = –80 nT according to the catalog of* Yermolaev *et al. falls into the Sheath interval.*

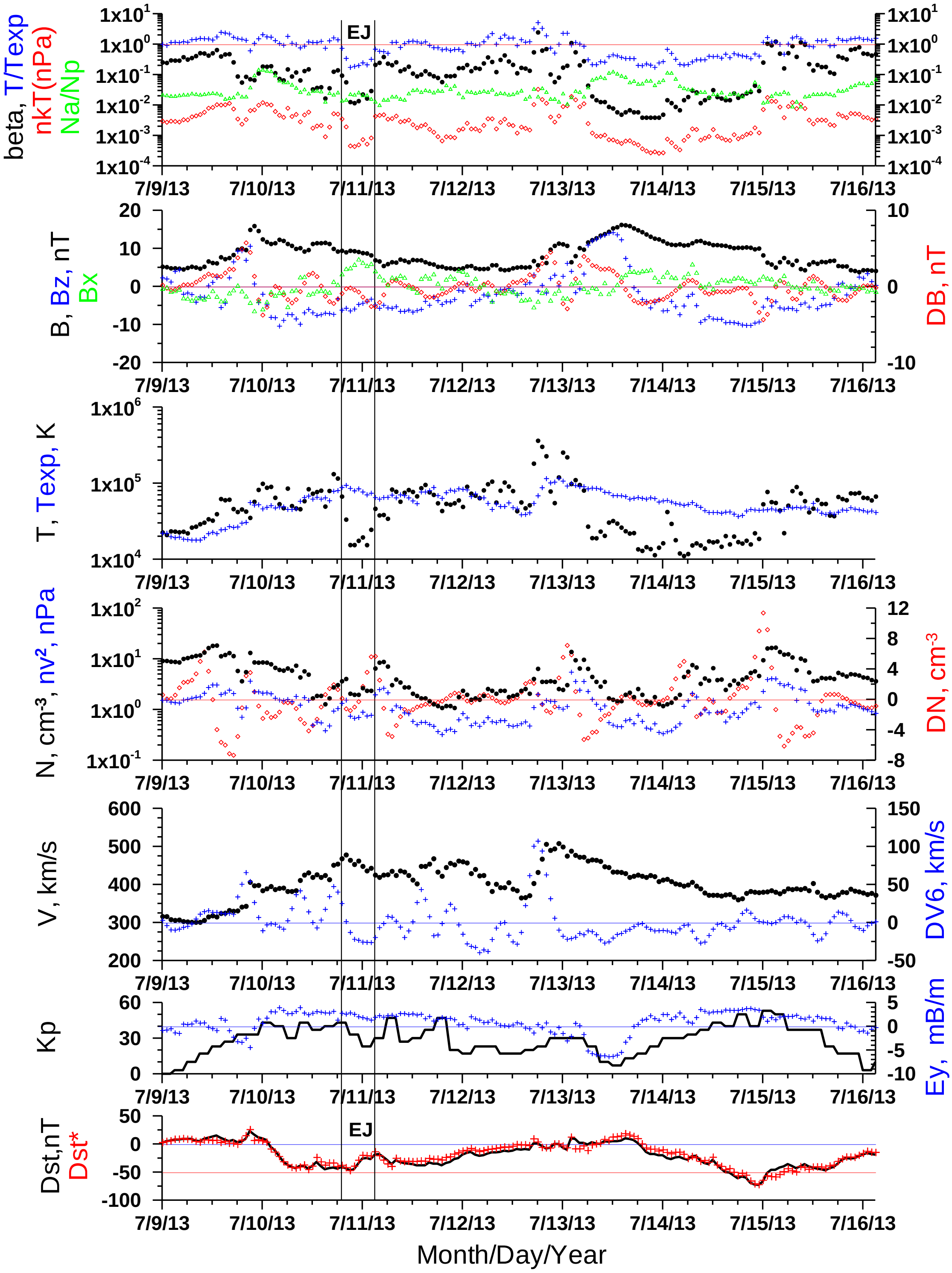

*Fig. 3. Same as fig. 1 Event 58 of the list by Qiu et al with a minimum of MS at 10.VII.2013 22.00 with Dst = –56 nT according to the catalog of* Yermolaev *et al. falls within the Ejecta interval.*

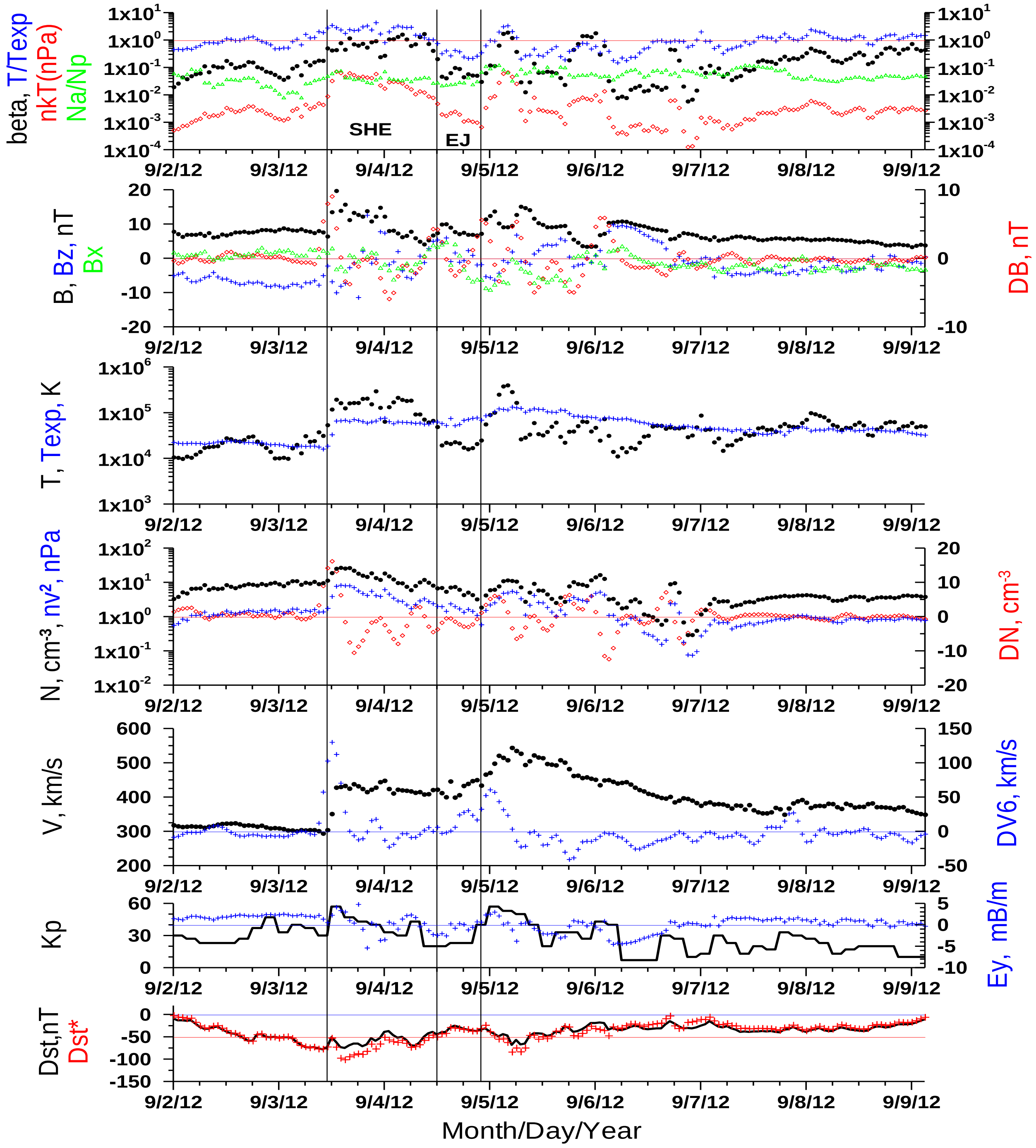

*Fig. 4 Same as fig. 1 Event 39 of the list by Qiu et al with a minimum of MS on 3.IX.2012 11.00 with Dst = –69 nT according to the catalog of* Yermolaev *et al. falls within the Sheath interval.*

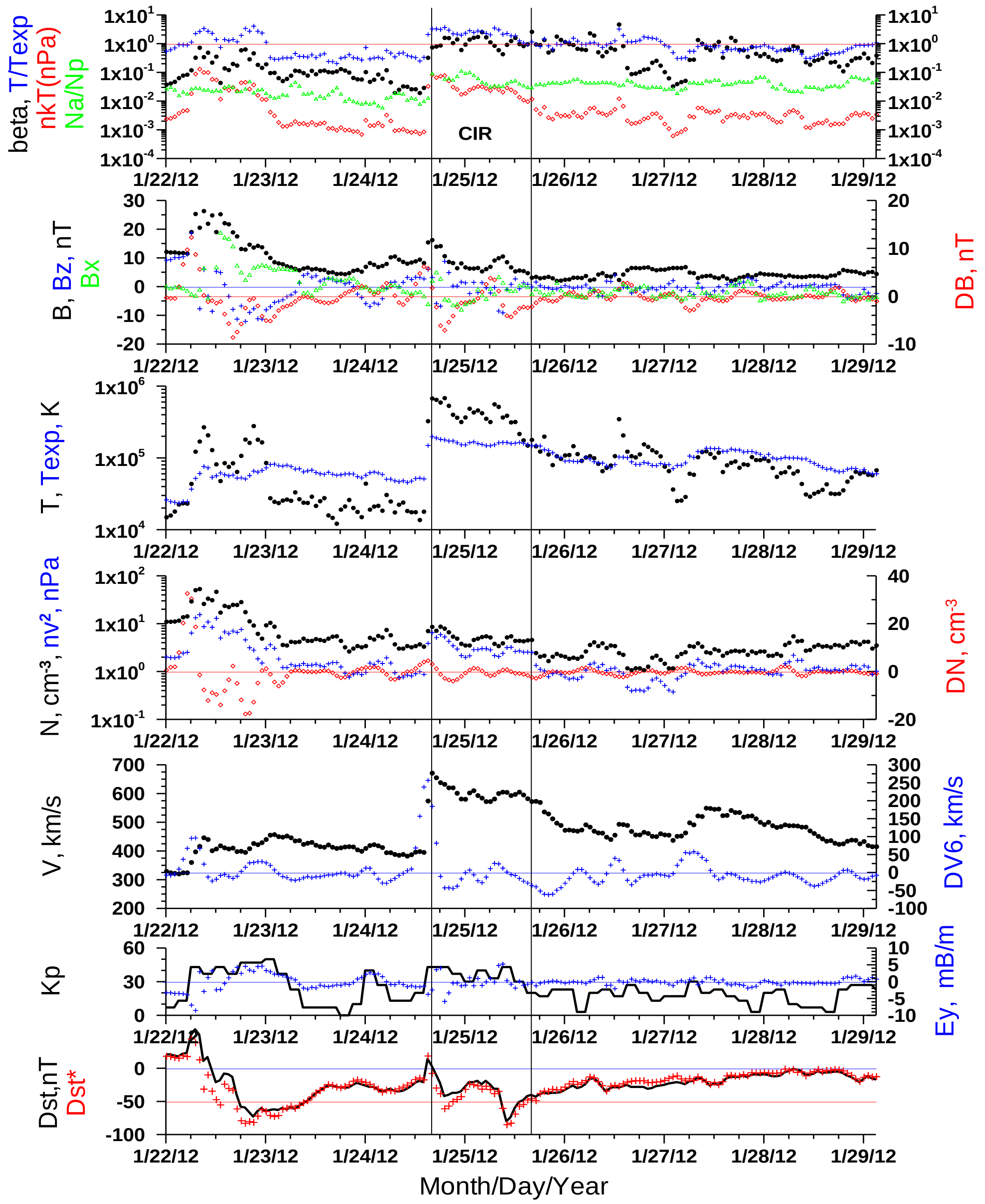

*Fig. 5. Same as fig. 1. Event 22 of the Qiu et al list with a minimum of MS on January 25, 2012 11.00 c Dst = –75 nT. According to the catalog by* Yermolaev *et al., it falls into the CIR interval.*

Table 1. Statistics comparing Yermolaev's data with Richardson and Cane.

| № | List of Qiu et al | Yermolaev et al. | | Catalog Richardson and Cane | |
|---|---|---|---|---|---|
| | Date, time, $Dst_{min}$, type CB | identifier Qiu et al | identifier | identifier Qiu et al | identifier Ермолаев |
| 1 | 22.VII.2009 07.00 –83 CIR | + | CIR | - | + |
| 2 | 16.II.2010 00.00 –59 ICME | - | SW | - | + |
| 3 | 6.IV.2010 16.00 –81 ICME | + | MCsh | + | + |
| 4 | 12.IV.2010 02.00 –67 SH | + | SH/Ejsh | + | + |
| 5 | 2.V.2010 19.00 –71 CIR | + | CIR | - | + |
| 6 | 29.V.2010 13.00 –80 ICME | + | MCsh | + | + |
| 7 | 4.VIII.2010 02.00 –74 SH | + | SH | + | + |
| 8 | 11.X.2010 20.00 –75 CIR | + | CIR | - | + |
| 9 | 4.II.2011 22.00 –63 CIR | + | CIR | - | + |
| 10 | 1.III.2011 15.00 –88 CIR | + | CIR | - | + |
| 11 | 11.III.2011 06.00 –83 CIR | - | EJsh | - | - |
| 12 | 6.IV.2011 20.00 –60 SH | - | SW | - | + |
| 13 | 28.V.2011 15.00 –80 ICME | + | MCsh | + | + |
| 14 | 5.VII.2011 01.00 –59 Complex | + | SW | - | + |
| 15 | 6.VIII.2011 04.00 –115 SH | + | CIR | + | - |
| 16 | 10.IX.2011 05.00 –75 CIR | - | MCsh | + | + |
| 17 | 17.IX.2011 16.00 –72 SH+ICME | + | MCsh | + | + |
| 18 | 27.IX.2011 00.00 –118 SH | + | MCsh | + | + |
| 19 | 25.X.2011 02.00 –147 SH+ICME | + | MCsh | + | + |
| 20 | 1.XI.2011 16.00 –66 ICME | - | SW | + | - |
| 21 | 22.I2012 22.00 –70 SH | + | SHmc | - | - |
| 22 | 25.I.2012 11.00 –75 SH | - | CIR | - | + |
| 23 | 15.II.2012 18.00 –67 SH+ICME | + | EJsh | + | + |
| 24 | 19.II.2012 05.00 –63 CIR | + | CIR | - | + |
| 25 | 27.II.2012 20.00 –57 SH+ICME | + | SHmc | + | + |
| 26 | 4.III.2012 03.00 –50 ICME | + | EJ | + | + |
| 27 | 7.III.2012 10.00 –88 CIR | + | CIR | - | + |
| 28 | 9.III.2012 09.00 –145 SH+ICME | + | MCsh | + | + |
| 29 | 12.III.2012 17.00 –64 SH | + | SH | - | + |
| 30 | 15.III.2012 21.00 –88 SH | + | SH | + | + |
| 31 | 28.III.2012 05.00 –68 CIR | + | CIR | - | + |
| 32 | 5.IV.2012 08.00 –64 ICME | - | SW | - | + |
| 33 | 13.IV.2012 06.00 –60 CIR | + | CIR | - | + |
| 34 | 24.IV.2012 05.00 –120 ICME | + | MCsh | + | + |
| 35 | 12.VI.2012 02.00 –67 ICME | + | EJ | - | - |
| 36 | 17.VI.2012 14.00 –86 ICME | + | MCsh | + | + |
| 37 | 9.VII.2012 13.00 –78 ICME | + | EJsh | + | + |
| 38 | 15.VII.2012 17.00 –139 SH+ICME | + | MCsh | + | + |
| 39 | 3.IX.2012 11.00 –69 ICME | - | SHej | + | + |
| 40 | 5.IX.2012 06.00 –64 Complex | + | SHej/EJ | + | + |
| 41 | 1.X.2012 05.00 –122 ICME | + | MCsh | + | + |
| 42 | 9.X.2012 09.00 –109 SH+ICME | + | MCsh | + | + |
| 43 | 13.X.2012 08.00 –90 ICME | + | EJsh | + | + |
| 44 | 1.XI.2012 21.00 –65 ICME | + | MCsh | + | + |
| 45 | 14.XI.2012 08.00 –108 ICME | - | SHej | + | + |
| 46 | 18.I.2013 01.00 –52 ICME | + | EJsh | + | + |
| 47 | 26.I.2013 23.00 –51 CIR | + | CIR | - | + |
| 48 | 1.III.2013 11.00 –55 CIR | + | CIR | - | + |
| 49 | 17.III.2013 21.00 –132 SH+ICME | + | EJsh | + | + |
| 50 | 29.III.2013 18.00 –59 CIR | + | CIR | - | + |
| 51 | 1.V.2013 19.00 –72 ICME | + | SW | + | + |
| 52 | 18.V.2013 04.00 –61 SH | + | SHej | - | - |
| 53 | 25.V.2013 07.00 –59 SH | + | SHej | - | - |
| 54 | 1.VI.2013 09.00 –124 CIR | + | CIR | - | + |
| 55 | 7.VI.2013 06.00 –78 ICME | + | MC | + | + |
| 56 | 29.VI.2013 07.00 –102 ICME | + | EJsh | + | + |
| 57 | 6.VII.2013 19.00 –87 ICME | + | EJsh | + | + |
| 58 | 10.VII.2013 22.00 –56 SH | - | EJ | - | - |
| 59 | 14.VII.2013 23.00 –81 ICME | + | EJsh | + | + |
| 60 | 5.VIII.2013 03.00 –50 CIR | + | CIR | - | + |
| 61 | 27.VIII.2013 22.00 –59 CIR | + | CIR | - | + |
| 62 | 2.X.2013 08.00 –72 SH | + | SHmc | + | + |
| 63 | 9.X.2013 02.00 –69 Unknown | - | SHej | + | + |
| 64 | 31.X.2013 00.00 –56 Complex | + | SW | - | + |
| 65 | 7.XI.2013 13.00 –50 SH | - | SW | - | + |
| 66 | 9.XI.2013 09.00 –80 CIR | - | SHej | + | + |
| 67 | 11.XI.2013 08.00 –68 SH | + | SHej | - | - |
| 68 | 8.XII.2013 09.00 –66 CIR | + | CIR | - | + |
| 69 | 19.II.2014 09.00 –119 ICME | + | EJsh | - | - |
| 70 | 20.II.2014 13.00 –95 SH | - | SW | - | + |
| 71 | 22.II.2014 02.00 –64 ICME | + | EJ | + | + |
| 72 | 23.II.2014 20.00 –55 Complex | + | sw | - | + |
| 73 | 28.II.2014 00.00 –97 SH | - | CIR | - | + |
| 74 | 12.IV.2014 10.00 –87 ICME | + | EJ | + | + |
| 75 | 30.IV.2014 10.00 –67 ICME | + | EJsh | + | + |
| 76 | 27.VIII.2014 19.00 –79 ICME | + | MC | - | - |
| 77 | 13.IX.2014 00.00 –88 ICME | + | MCsh | + | + |
| 78 | 9.X.2014 08.00 –51 Complex | + | SW | - | + |
| 79 | 28.X.2014 02.00 –57 Unknown | + | SW | - | + |
| 80 | 10.XI.2014 18.00 –65 Unknown | + | SW | - | + |
| 81 | 16.XI.2014 08.00 –59 Unknown | + | SW | - | + |
| 82 | 12.XII.2014 17.00 –53 Unknown | - | CIR | - | + |
| 83 | 22.XII.2014 07.00 –71 ICME | + | MCsh | + | + |
| 84 | 24.XII.2014 01.00 –57 SH | - | SW | - | + |
| 85 | 4.I.2015 22.00 –71 ICME | - | CIR | - | + |
| 86 | 7.I.2015 12.00 –99 ICME | + | MC | + | + |
| 87 | 18.II.2015 01.00 –64 CIR | + | CIR | - | + |
| 88 | 8.II.2015 08.00 –56 CIR | + | CIR | - | + |
| 89 | 2.III.2015 09.00 –55 CIR | - | SHej | - | - |
| 90 | 17.III.2015 23.00 –222 SH+ICME | + | MCsh | + | + |
| 91 | 11.IV.2015 11.00 –75 ICME | + | MCsh | + | + |
| 92 | 17.IV.2015 00.00 –79 CIR | + | CIR | - | + |
| 93 | 11.V.2015 05.00 –51 ICME | + | MCsh | + | + |
| 94 | 13.V.2015 07.00 –76 CIR | + | CIR | - | + |
| 95 | 8.VI.2015 09.00 –73 CIR | + | CIR | - | + |
| 96 | 23.VI.2015 05.00 –204 SH+ICME | + | MCsh | + | + |
| 97 | 25.VI.2015 20.00 –86 ICME | + | MCsh | + | + |
| 98 | 7.V.2015 06.00 –67 CIR | + | CIR | - | + |
| 99 | 13.VII.2015 16.00 –61 ICME | + | EJsh | + | + |
| 100 | 23.VII.2015 09.00 –63 Complex | - | CIR | - | + |
| 101 | 16.VIII.2015 08.00 –84 SH+ICME | + | MCsh | + | + |
| 102 | 19.VIII.2015 07.00 –50 Complex | + | SW | - | + |
| 103 | 27.VIII.2015 21.00 –92 ICME | + | EJsh | + | + |
| 104 | 7.IX.2015 21.00 –70 SH | + | SH/MCsh | - | - |
| 105 | 9.IX.2015 13.00 –98 ICME | + | MCsh | + | + |
| 106 | 11.IX.2015 15.00 –81 CIR | + | CIR | - | + |
| 107 | 20.IX.2015 16.00 –75 SH | + | SHej | - | - |
| 108 | 7.X.2015 23.00 –124 CIR | + | CIR | - | + |
| 109 | 4.XI.2015 13.00 –60 CIR | - | EJsh | + | + |
| 110 | 7.XI.2015 07.00 –89 SH | + | SH/MCsh | + | + |
| 111 | 10.XI.2015 14.00 –58 CIR | + | CIR | - | + |
| 112 | 20.XII.2015 23.00 –155 ICME | + | MCsh | + | + |
| 113 | 1.I.2016 01.00 –110 SH+ICME | + | EJsh | - | - |
| 114 | 20.I.2016 17.00 –93 ICME | - | SW | + | - |
| 115 | 3.II.2016 03.00 –53 CIR | - | EJsh | - | - |
| 116 | 16.II.2016 20.00 –57 CIR | + | CIR | - | + |
| 117 | 6.III.2016 22.00 –98 CIR | + | CIR | - | + |
| 118 | 3.IV.2016 00.00 –56 CIR | + | CIR | - | + |
| 119 | 8.IV.2016 01.00 –60 ICME | + | EJsh | - | - |
| 120 | 13.IV.2016 06.00 –55 CIR | + | CIR | - | + |
| 121 | 14.IV.2016 21.00 –59 ICME | + | EJ | + | + |
| 122 | 8.IV.2016 09.00 –88 CIR | + | CIR | - | + |
| 123 | 3.VIII.2016 11.00 –52 SH | - | CIR2 | - | + |
| 124 | 23.VIII.2016 22.00 –74 CIR | + | CIR | - | + |
| 125 | 1.IX.2016 10.00 –59 CIR | + | CIR | - | + |
| 126 | 14.X.2016 00.00 –104 ICME | + | MCsh | + | + |
| 127 | 25.X.2016 18.00 –59 CIR | + | CIR | - | + |
| 128 | 29.X.2016 04.00 –64 CIR | + | CIR | - | + |
| 129 | 1.III.2017 22.00 –61 CIR | + | CIR | - | + |
| 130 | 27.III.2017 15.00 –74 CIR | + | CIR | - | + |
| 131 | 22.IV.2017 17.00 –50 CIR | + | CIR/ SW | - | + |
| 132 | 28.V.2017 08.00 –125 ICME | + | MCsh | + | + |
| 133 | 16.VII.2017 16.00 –72 SH | + | SHmc | + | + |
| 134 | 31.VIII.2017 12.00 –50 CIR | - | EJsh | - | - |
| 135 | 8.IX.2017 02.00 –124 ICME | - | SHmc | + | + |
| 136 | 28.IX.2017 07.00 –55 CIR | + | CIR | - | + |
| 137 | 14.X.2017 06.00 –52 CIR | - | SW | - | + |
| 138 | 8.XI.2017 02.00 –74 CIR | + | CIR | - | + |
| 139 | 18.III.2018 22.00 –50 CIR | + | CIR | - | + |
| 140 | 20.IV.2018 10.00 –66 CIR | + | CIR | - | + |
| 141 | 6.V.2018 02.00 –56 CIR | + | CIR | - | + |
| 142 | 26.VIII.2018 07.00 –174 ICME | + | MCsh | + | + |
| 143 | 11.IX.2018 11.00 –60 CIR | - | SHej | - | - |
| 144 | 7.X.2018 22.00 –53 CIR | + | CIR | - | + |
| 145 | 5.XI.2018 06.00 –53 CIR | + | CIR | - | + |
| 146 | 11.V.2019 22.00 –51 SH+ICME | + | MC | + | + |
| 147 | 14.V.2019 08.00 –65 SH+ICME | + | SHmc | + | + |
| 148 | 5.VIII.2019 21.00 –53 CIR | + | CIR | - | + |
| 149 | 1.IX.2019 07.00 –52 CIR | - | SW | - | + |

* Column contents Table.1:

1- column – event numbering

2- column – date and time of events (minimum MS) in day-month-year hour.min format (UT), Dst index value at minimum MS, type CB in minimum MS list Qiu et al (ICME - interplanetary coronal mass ejections (Ejecta, MC), SH – shell Sheath, SH+ICME - combination of shells and interplanetary coronal mass ejections, CIR - corotating interaction areas, Complex- сложные конструкции, Unknown (Unknown) - missing data and poor data quality.

3- column – type match identifier CB on the list Qiu et al al with catalog Yermolaev et al. : (“+“ или “-“ means a match or non-match of the event).

4- column - type SW according to the catalog of Yermolaev et al. at a minimum MS, the time of which is indicated in the 2nd column.

5- column – presence of an event in the list Richardson and Cane (“+“ или “-“means a match or non-match of the event ICME).

6- column – catalog event match Yermolaev et al. and catalog Richardson and Cane. Since the catalog R were previously “coarsened” to catalog nomenclature ichardson and Cane contains only events “ICME“, then the data to the catalog Yermolaev et al. were previously “coarsened” to catalog nomenclature Richardson and Cane, included only “ICME“ amd “no-ICME“

Table 2. Links to figures in the catalog by Yermolaev et al., for events that have differences in the identification of the list by Qiu et al from Table 1.

| № | List of events Qiu et al | Type CB | ftp://ftp.iki.rssi.ru/pub/ omni/catalog/ | ftp://ftp.iki.rssi.ru/pub/ omni/ |
|---|---|---|---|---|
| 11 | 11.III.2011 06.00 –83 CIR | EJsh | 2011/20110226c.jpg | 2011/20110305jpg |
| 16 | 10.IX.2011 05.00 –75 CIR | MCsh | 2011/20110910c.jpg | 2011/20110903.jpg 2011/20110910.jpg |
| 109 | 4.XI.2015 13.00 –60 CIR | EJsh | 2015/20151008c.jpg | 2015/20151029.jpg |
| 115 | 3.II.2016 03.00 –53 CIR | EJsh | 2016/20160129c.jpg | 2016/20160129.jpg |
| 134 | 31.VIII.2017 12.00 –50 CIR | EJsh | 2017/20170813c.jpg | 2017/20170827.jpg |
| 66 | 9.XI.2013 09.00 –80 CIR | SHej | 2013/20131105c.jpg | 2013/20111053.jpg |
| 89 | 3.VI.2015 09.00 –55 CIR | SHej | 2015/20150226c.jpg | 2015/20150226.jpg |
| 143 | 11.IX.2018 11.00 –60 CIR | SHej | 2018/20180910c.jpg | 2018/20180910.jpg |
| 137 | 14.X.2017 06.00 –52 CIR | SW | 2017/20171008c.jpg | 2017/20171008.jpg |
| 149 | 1.IX.2019 07.00 –52 CIR | SW | 2019/20190813c.jpg | 2019/20190827.jpg |
| 58 | 10.VII.2013 22.00 –56 SH | EJ | 2013/20130618c.jpg | 2013/20130709c.jpg |
| 39 | 3.IX.2012 11.00 –69 ICME | SHej | 2012/20120812c.jpg | 2012/20120902.jpg |
| 45 | 14.XI.2012 08.00 –108 ICME | SHej | 2012/20121108c.jpg | 2012/20121111.jpg |
| 63 | 9.X.2013 02.00 –69 Unknown | SHej | 2013/20131008c.jpg | 2013/20131008.jpg |
| 135 | 8.IX.2017 02.00 –124 ICME | SHej | 2017/20170715c.jpg | 2017/20170903.jpg |
| 22 | 25.I.2012 11.00 –75 SH | CIR | 2012/20120101c.jpg | 2012/20120122.jpg |
| 73 | 28.II.2014 00.00 –97 SH | CIR | 2014/20140226c.jpg | 2014/20140226.jpg |
| 82 | 12.XII.2014 17.00 –53Unknown | CIR | 2014/20141203c.jpg | 2014/20141210.jpg |
| 85 | 4.I.2015 22.00 –71 ICME | CIR | 2015/20150101c.jpg | 2015/20150101.jpg |
| 100 | 23.VII.2015 09.00 –63 Complex | CIR | 2015/20150716c.jpg | 2015/20150723.jpg |
| 123 | 3.VIII.2016 11.00 –52 SH | CIR | 2016/20160715c.jpg | 2016/20160729.jpg |
| 2 | 16.II.2010 00.00 –59 ICME | SW | 2010/20100129c.jpg | 2010/20100216.jpg |
| 12 | 6.IV.2011 20.00 –60 SH | SW | 2011/20110326c.jpg | 2011/20110402.jpg |
| 20 | 1.XI.2011 16.00 –66 ICME | SW | 2011/20111008c.jpg | 2011/20111029.jpg |
| 32 | 5.IV.2012 08.00 –64 ICME | SW | 2012/20120325c.jpg | 2012/20120401.jpg |
| 65 | 7.XI.2013 13.00 –50 SH | SW | 2013/20131105c.jpg | 2013/20131105.jpg |
| 70 | 20.II.2014 13.00 –95 SH | SW | 2014/20140129c.jpg | 2014/20140219.jpg |
| 84 | 24.XII.2014 01.00 –57 SH | SW | 2014/20141203c.jpg | 2014/20141224.jpg |
| 114 | 20.I.2016 17.00 –93 ICME | SW | 2016/20160101c.jpg | 2016/20160115.jpg |

* Column contents Table.1:

1, 2 column – similarly Table. 1

3- column – type CB events according to the catalog Yermolaev et al.

4, 5- column - name of the drawing file in the catalog Yermolaev et al.